\documentclass{egpubl}

\Paper
\usepackage[T1]{fontenc}
\usepackage{dfadobe}  

\usepackage{microtype}
\usepackage{newtxtext}
\usepackage[slantedGreek]{newtxmath}
\usepackage{mathtools}
\def\clap#1{\hbox to 0pt{\hss #1\hss}}%

\usepackage{cite}  %
\BibtexOrBiblatex
\electronicVersion
\PrintedOrElectronic
\ifpdf \usepackage[pdftex]{graphicx} \pdfcompresslevel=9
\else \usepackage[dvips]{graphicx} \fi

\usepackage{egweblnk} 

\usepackage{amsfonts}
\usepackage{amsmath}
\usepackage{array}
\usepackage{booktabs}
\usepackage[capitalize]{cleveref}
\crefname{paragraph}{Sec.}{Secs.}
\crefname{section}{Sec.}{Secs.}
\Crefname{section}{Section}{Sections}
\Crefname{table}{Table}{Tables}
\crefname{table}{Tab.}{Tabs.}
\Crefname{figure}{Figure}{Figures}
\crefname{figure}{Fig.}{Figs.}
\Crefname{chapter}{Chapter}{Chapters}
\crefname{chapter}{Ch.}{Chs.}

\def\vect#1{\boldsymbol{#1}}
\def\topVert{\vphantom{\vert}\raisebox{-0.5ex}{$\vert$}}
\def\botVert{\vphantom{\vert}\smash{\raisebox{0.5ex}{$\vert$}}}

\usepackage{multicol, multirow}
\usepackage{xcolor}
\usepackage{colortbl}
\definecolor{1st}{RGB}{102,194,164} %
\definecolor{2nd}{RGB}{178,226,226} %
\definecolor{3rd}{RGB}{237,248,251} %
\definecolor{1stText}{RGB}{57,146,116} %
\newcommand\tstrut{\rule{0pt}{2.4ex}}
\newcommand\bstrut{\rule[-1.0ex]{0pt}{0pt}}
\usepackage{arydshln}
\title[Efficient Perspective-Correct 3D Gaussian Splatting Using Hybrid Transparency]%
      {Efficient Perspective-Correct 3D Gaussian Splatting\\Using Hybrid Transparency}

\author[F. Hahlbohm et al.]
{\parbox{\textwidth}{\centering%
    Florian Hahlbohm$^{1}$\hspace{2pt}\orcid{0009-0004-8710-1433}\hspace{10pt}
    Fabian Friederichs$^{1}$\hspace{2pt}\orcid{0000-0003-0777-7229}\hspace{10pt}
    Tim Weyrich$^{2,3}$\hspace{2pt}\orcid{0000-0002-4322-8844}\hspace{10pt}
    Linus Franke$^{2}$\hspace{2pt}\orcid{0000-0001-8180-0963}\hspace{10pt}
    Moritz Kappel$^{1}$\hspace{2pt}\orcid{0000-0001-9507-5141}\\
    Susana Castillo$^{1}$\hspace{2pt}\orcid{0000-0003-1245-4758}\hspace{10pt}
    Marc Stamminger$^{2}$\hspace{2pt}\orcid{0000-0001-8699-3442}\hspace{10pt}
    Martin Eisemann$^{1}$\hspace{2pt}\orcid{0000-0002-8673-4405}\hspace{10pt}
    Marcus Magnor$^{1,4}$\hspace{2pt}\orcid{0000-0003-0579-480X}
}\\
\parbox{\textwidth}{\centering%
$^1$Computer Graphics Lab, TU Braunschweig, Germany\hspace{7pt}\texttt{\{lastname\}@cg.cs.tu-bs.de}\\
$^2$ Visual Computing Erlangen, FAU Erlangen-Nürnberg, Germany\hspace{7pt}\texttt{\{firstname.lastname\}@fau.de}\\
$^3$University College London, United Kingdom\hspace{7pt}$^4$University of New Mexico, USA\\
\vspace{15pt}
\url{https://fhahlbohm.github.io/htgs/}
}
}

\begin{document}

\teaser{
\vspace{-15pt}
\includegraphics[width=0.9\linewidth]{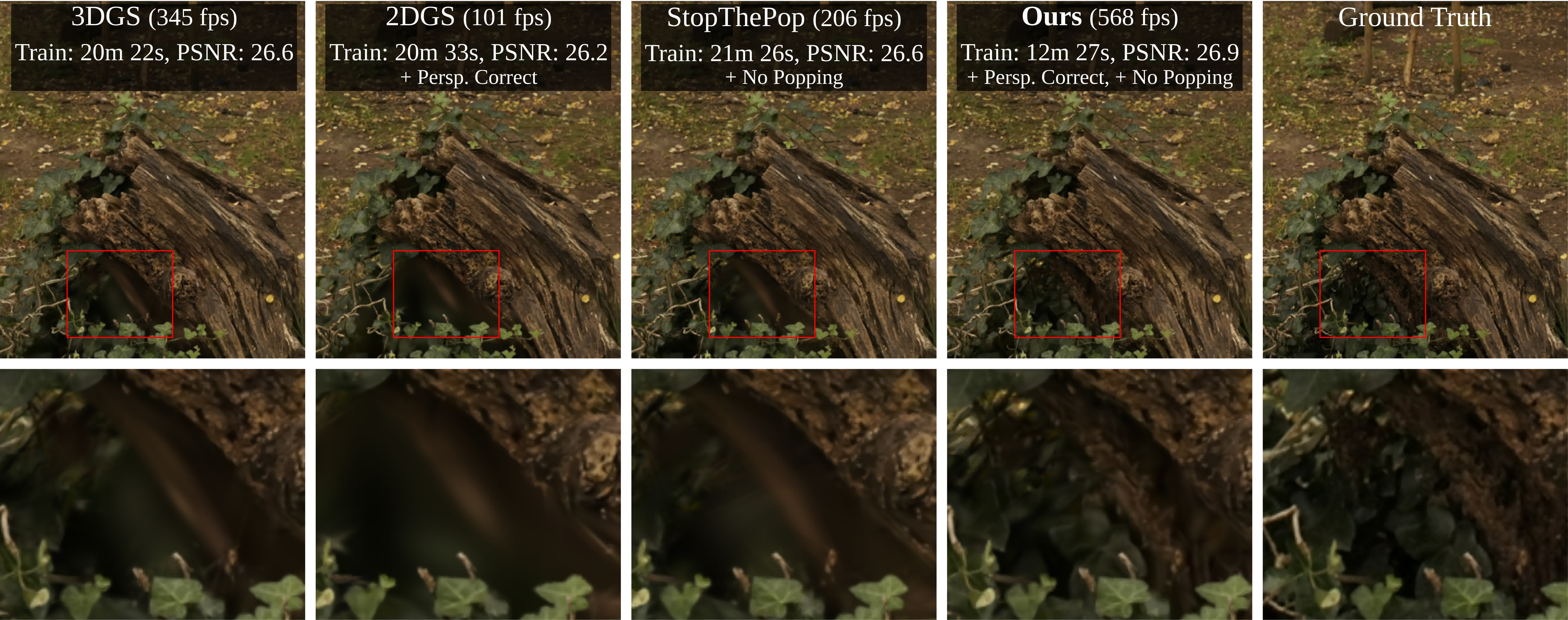} 
\centering
\caption{%
We present the first perspectively and simultaneously geometrically accurate approach for real-time rendering of 3D Gaussian splats. Our temporally stable blending formulation based on hybrid transparency effectively removes the popping artifacts caused by the approximate, global sorting scheme of 3D Gaussian Splatting (3DGS)~\cite{kerbl3Dgaussians}. Additionally, we replace the approximate projection used in 3DGS with a novel, numerically stable formulation for evaluation of general 3D Gaussians along per-pixel rays to improve rendering quality. We also achieve a significant speedup compared to state-of-the-art approaches addressing these challenges~\cite{huang20242d, radl2024stopthepop}, as our blending formulation only requires partial depth-ordering and our splat evaluation is well-suited for rasterization.}\label{fig:teaser}
}

\maketitle

\begin{abstract}
3D Gaussian Splats (3DGS) have proven a versatile rendering primitive, both for inverse rendering as well as real-time exploration of scenes. In these applications, coherence across camera frames and multiple views is crucial, be it for robust convergence of a scene reconstruction or for artifact-free fly-throughs. Recent work started mitigating artifacts that break multi-view coherence, including popping artifacts due to inconsistent transparency sorting and perspective-correct outlines of (2D) splats. At the same time, real-time requirements forced such implementations to accept compromises in how transparency of large assemblies of 3D Gaussians is resolved, in turn breaking coherence in other ways. In our work, we aim at achieving maximum coherence, by rendering fully perspective-correct 3D Gaussians while using a high-quality approximation of accurate blending, hybrid transparency, on a per-pixel level, in order to retain real-time frame rates.
Our fast and perspectively accurate approach for evaluation of 3D Gaussians does not require matrix inversions, thereby ensuring numerical stability and eliminating the need for special handling of degenerate splats, and the hybrid transparency formulation for blending maintains similar quality as fully resolved per-pixel transparencies at a fraction of the rendering costs. We further show that each of these two components can be independently integrated into Gaussian splatting systems.
In combination, they achieve up to {2\texttimes} higher frame rates, {2\texttimes} faster optimization, and equal or better image quality with fewer rendering artifacts compared to traditional 3DGS on common benchmarks.

\ccsdesc[500]{Computing methodologies~Rendering}
\ccsdesc[500]{Computing methodologies~Point-based models}
\ccsdesc[500]{Computing methodologies~Rasterization}
\ccsdesc[500]{Computing methodologies~Machine learning approaches}
\printccsdesc

\end{abstract}

\section{Introduction}
\label{sec:intro}

Novel view synthesis has undergone a significant transformation with the advent of Neural Radiance Fields (NeRF)~\cite{mildenhall2020nerf} and 3D Gaussian Splatting (3DGS)~\cite{kerbl3Dgaussians}, both of which have established themselves as powerful techniques for rendering complex 3D scenes. 
Among these, 3DGS has emerged as the de-facto representation for user-facing radiance field applications due to its fast optimization and real-time rendering capabilities.
Its speed and efficiency stem from an explicit point-based representation, where each point's extent is modeled by an anisotropic 3D Gaussian. 
Using fast, tile-based rasterization, this can be efficiently implemented on modern GPUs.

However, the remarkable performance of 3DGS is achieved through a series of approximations that, while effective in many scenarios, introduce limitations in multi-view consistency.
One significant issue arises from the affine approximation used in projecting 3D Gaussian splats onto the image plane, causing differing pixel contributions depending on camera placement. 
Another one is the per-primitive sorting, which causes elliptical color patches to suddenly appear or disappear during movement (called popping~\cite{radl2024stopthepop}).
In order to solve the issues, it is necessary to accurately sort and evaluate Gaussians per pixel, introducing a strong bottleneck on the sorting stage.
In this paper, we introduce an pipeline which improves the original 3DGS framework in eliminating these artifacts while offering superior performance through a hybrid transparency approximation.

\begin{figure}[b!]
\centering
\begin{tabular}{@{}p{0.33333\linewidth}<{\centering}@{}p{0.33333\linewidth}<{\centering}@{}p{0.33333\linewidth}<{\centering}@{}}
 & 3DGS & Ours
\end{tabular}
\includegraphics[width=\linewidth]{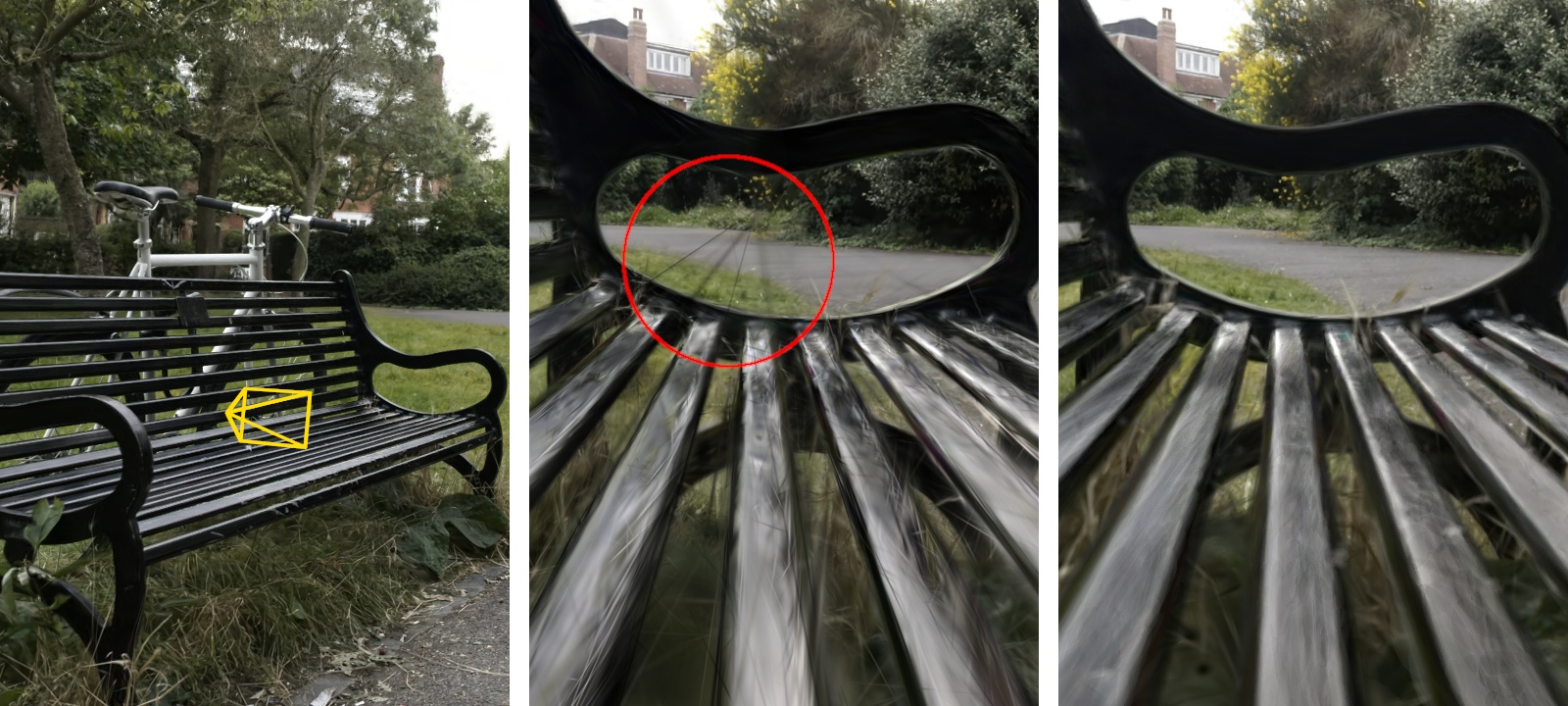}
\caption{%
In 3DGS~\cite{kerbl3Dgaussians}, Kerbl et al. use an affine approximation for projecting 3D Gaussians. Common benchmarks do not include viewpoints where this approximation matter is highlighted. Here, we demonstrate the benefits of our perspective-correct projection by comparing renderings from 3DGS and our method.
}\label{fig:persp_correct}
\end{figure}

For projection artifacts, although the affine approximation performs well on benchmark datasets, it fails to model perspective distortion correctly, especially when parts of the scene are viewed at close distances.
The result are visually disturbing artifacts, where the projected Gaussians take on extreme, distorted shapes, severely affecting the rendering quality (see \cref{fig:persp_correct}).

Recent work on 2D Gaussian Surfels~\cite{huang20242d} has made strides in achieving perspective-accurate rendering by leveraging established techniques from Sigg~et~al.~\cite{sigg2006quadrics} and Weyrich~et~al.~\cite{weyrich2007hardware}.
For 3D Gaussians, concurrent research explores evaluating splats by calculating intersection points with per-pixel viewing rays~\cite{yu2024gof}.
This approach can be implemented in ray-tracing frameworks~\cite{moenne2024gaussiantracer}, whose rendering speeds are significantly lower. 
Alternatively, the rasterization-based approach faces numerical instability due to the matrix inversions required, making optimization via gradient descent challenging and prone to catastrophic failure if handled incorrectly.

In this paper, we propose a fast, differentiable method for perspective-accurate 3D Gaussian splat evaluation at the point of maximum contribution along per-pixel viewing rays that avoids matrix inversion entirely.

The second issue with 3DGS lies in depth ordering during rendering, where only the view-space depth of the mean is considered. 
This causes incorrect per-pixel blending order of fragments, leading to the so-called ``popping'' artifacts that disrupt multi-view consistency and especially the viewing experience in motion. 
Recent work, such as \textit{StopThePop}~\cite{radl2024stopthepop}, addresses this through hierarchical sorting, which is comprised of global presorting and then locally sorting within a
sliding window, with progressively lower tile sizes.

\begin{figure}[bht]
\centering
\begin{tabular}{@{}p{0.39855072463\linewidth}<{\centering}@{}p{0.40579710144\linewidth}<{\centering}@{}p{0.19565217391\linewidth}<{\centering}@{}}
View & Rotated View & 
\end{tabular}
\includegraphics[width=\linewidth]{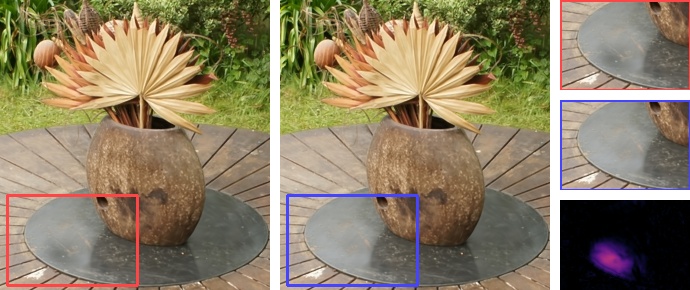}
\begin{tabular}{@{}p{\linewidth}<{\centering}@{}}
Approximate Sorting $+$ Alpha Blending (3DGS)\\
\hdashline
\end{tabular}
\begin{tabular}{@{}p{0.39855072463\linewidth}<{\centering}@{}p{0.40579710144\linewidth}<{\centering}@{}p{0.19565217391\linewidth}<{\centering}@{}}
View & Rotated View & 
\end{tabular}
\includegraphics[width=\linewidth]{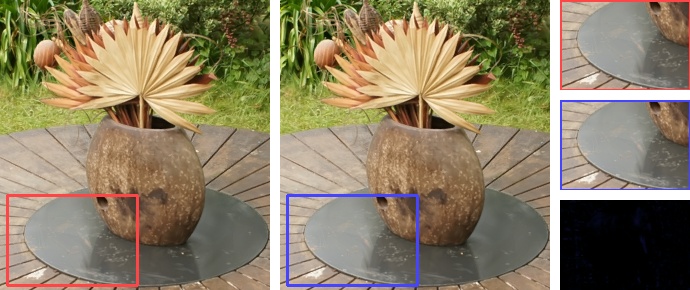}
\begin{tabular}{@{}p{\linewidth}<{\centering}@{}}
Hybrid Transparency (Ours)
\end{tabular}
\caption{%
Radl et al.~\cite{radl2024stopthepop} discuss how 3DGS approximates depth-ordering for alpha blending, leading to popping artifacts during view rotation. To solve this without sacrificing performance, we propose a hybrid transparency approach, combining alpha blending with order-independent transparency, resulting in temporally stable rendering and an improved viewing experience.
}\label{fig:popping}
\end{figure}

We propose a similar idea, based on the established rendering paradigm of \textit{Hybrid Transparency}~\cite{Wyman2016}, which lets us skip the global presorting and provides high performance.
By alpha-blending the first $K$ fragments (called the \textit{core}) in correct depth-order per pixel and accumulating remaining contributions (the \textit{tail}) using an order-independent residual, our method mitigates the popping artifacts (see \cref{fig:popping}) while maintaining superior performance.

These two enhancements -- perspective-accurate splat evaluation and improved depth-ordering -- can be integrated into existing 3DGS systems independently.
In our implementation, we combine both improvements, demonstrating that they not only eliminate the respective artifacts but also enhance rendering speed, crucial for real-time applications that rely on fast scene inspection.
In summary, we make the following contributions:

\begin{itemize}
\setlength\itemsep{.5pt}
\item We present a novel, matrix inversion-free ray-splat intersection method within a differentiable renderer for 3D Gaussian splats, resulting in numerically stable optimization and fully perspective-correct rendering.
\item We introduce hybrid transparency to 3D Gaussian splat rendering, resulting in both very stable results and improved performance.
\item We evaluate our efficient implementation, which outperforms established methods significantly, while matching or exceeding their visual fidelity.
\end{itemize}

\section{Related Work}
\label{sec:related}
This work is at the intersection of novel view synthesis, Gaussian splatting, and order-independent transparency.

\subsection{Novel View Synthesis}
Novel view synthesis aims to render images of a scene from arbitrary viewpoints given a fixed number of input images.

The field has recently been revolutionized by Neural Radiance Fields (NeRF) introduced by Mildenhall~et~al.~\cite{mildenhall2020nerf}.
NeRF represents a scene by optimizing a large MLP that maps positions and viewing directions to volumetric density and color. 
The use of differentiable volume rendering enables optimization using gradient descent, which leads to high-quality results; however, NeRF's slow training and rendering process hinders its use in interactive applications.
To accelerate NeRF, voxel grids with trilinear interpolation were introduced, offering faster, continuous representations~\cite{yu2021plenoxels, sun2022dvgo}. 
Memory-efficient methods, such as multi-resolution hash tables~\cite{mueller2022instant} and tensor factorization~\cite{Chen2022tensorf, reiser2023merf}, further reduce computational overhead.
The current state-of-the-art method in terms of image quality, Zip-NeRF~\cite{barron2023ICCV}, combines these grid-based methods with solutions addressing aliasing issues~\cite{barron2021mipnerf, barron2022mipnerf360}.

Alternatively, point-based models use explicit point clouds for geometry, rendering images through fast rasterization. 
Due to the discrete nature of point clouds, recent point-based radiance field methods employ large convolutional neural networks (CNN) for hole-filling in image-space~\cite{aliev2020npbg, ruckert2022adop, franke2023vet, kopanas2021perviewopt,harrerfranke2023inovis} or MLPs for feature decoding~\cite{kopanas2022catacaustics}, but these NNs are computationally expensive. 
Addressing this limitation, Franke~et~al.\ showed that assigning a radius to each point in combination with trilinear interpolation into an image pyramid allows using a smaller CNN~\cite{franke2024trips}.
While enabling fast rendering, image quality of point-based models is often lacking behind implicit NeRF- and grid-based models.
Very recently, Hahlbohm~et~al.\ showed that point-based models and grid-based models can be combined to improve image quality and robustness~\cite{hahlbohm2024inpc}.
Remaining issues with point-based models are temporal instabilities and the reliance on dedicated GPUs, which are both due to the used CNN.
Another approach involves converting a trained NeRF model into a faster representation for inference~\cite{hedman2021snerg, yu2021plenoctrees}.
Recent methods achieve impressive frame rates with only a small quality loss via triangle mesh baking~\cite{yariv2023bakedsdf, reiser2024binary} or distilling of state-of-the-art NeRF models into memory-efficient, hybrid representations~\cite{duckworth2023smerf}.
While they render exceptionally fast even on low-end mobile devices, a major disadvantage is their reliance on the NeRF model that needs to be trained before the -- often equally expensive -- baking process.
Other approaches that proved effective directly optimize a triangle mesh~\cite{chen2023mobilenerf} or a set of tetrahedra~\cite{kulhanek2023tetra}.

\subsection{Gaussian Splatting}
In early work on point-based rendering, points were rasterized as opaque splats~\cite{grossman1998point,rusinkiewicz2000qsplat,pfister2000surfels} and thus suffered from strong aliasing.
It was shown that quality can be significantly improved using semitransparent splats~\cite{zwicker2001surfsplat}, however this requires costly sorting and blending operations.
To address performance issues, fast algorithms for then GPUs~\cite{botsch2005high} or dedicated hardware implementations~\cite{weyrich2007hardware} were proposed.

More recently, point-based rendering regained momentum as a powerful primitive for differentiable rendering in novel view synthesis~\cite{kopanas2021perviewopt,aliev2020npbg,ruckert2022adop}.
In 2023, Kerbl~et~al.~\cite{kerbl3Dgaussians} re-introduced rasterization-based rendering of point primitives with anisotropic Gaussian extent in their seminal work \textit{3D Gaussian Splatting (3DGS)}.
Utilizing fast, tile-based rasterization~\cite{lassner2021pulsar}, 3DGS optimizes a set of explicit 3D Gaussians via gradient descent and an adaptive density control mechanism.
Due exceptionally fast rendering during inference, 3DGS has inspired a plethora of follow-up research addressing, e.g., anti-aliasing~\cite{yu2024mip}, compression~\cite{3DGSzip2024}, dynamic scenes~\cite{luiten2023dynamic, wu20244dgs}, large scenes~\cite{kerbl2024hierarchical}, and the reliance on an initial point cloud~\cite{kheradmand243dgsmcmc, niemeyer2024radsplat}.
Especially relevant to this work is \textit{StopThePop} by Radl~et~al.~\cite{radl2024stopthepop}, which employs a hierarchical sorting approach to reduce popping artifacts originating from approximate sorting used in 3DGS.
Notably, the per-pixel ordering resulting from this hierarchical sorting approach is not guaranteed to be fully correct.
We also present a method for avoiding these popping artifacts, but achieve this by changing the blending procedure to a hybrid transparency-based approach so that our method only requires partial depth-ordering of the initial $K$ splats contributing to each pixel color.
Thus, we avoid the need for a sophisticated sorting implementation which, even when compared to the approximate solution used by 3DGS, improves performance.
Equally important in the context of this work is the perspectively accurate \textit{2D Gaussian Splatting (2DGS)} by Huang~et~al.~\cite{huang20242d} as well as concurrent work that introduces ray tracing into the 3DGS framework~\cite{yu2024gof, moenne2024gaussiantracer}.
While the latter also provide a perspectively accurate rendering formulation by using ray tracing, they introduce a significant slowdown during both the optimization and subsequent inference.
Furthermore, their evaluation of the 3D Gaussian primitives relies on matrix inversion, which is prone to numerical instabilities when Gaussians fall flat along one or more principal axes.
The perspectively correct approach by Huang~et~al.~\cite{huang20242d} circumvents these numerical instabilities by utilizing an optimized approach for 2D, i.e., degenerate 3D Gaussians~\cite{sigg2006quadrics, weyrich2007hardware}.
Regarding the perspectively correct rendering of non-degenerate 3D Gaussians, projecting each Gaussian onto a tangential plane of the unit sphere around the camera~\cite{huang2024erroranalysis} works well but significantly slows down optimization and subsequent inference.
In this work, we present an approach that enables perspectively correct rendering of 3D Gaussians through ray-splat intersection: we extend the perspectively correct approach of Weyrich~et~al.~\cite{weyrich2007hardware} to work with non-degenerate 3D Gaussians, which enables the use of ray-splat intersection for rasterization-based rendering with negligible overhead and no issues with respect to numerical instabilities.

\subsection{Order-Independent Transparency}
The seminal Porter-Duff algorithm~\cite{porterduff84} blends two semi-transparent surfaces correctly, but extending it to more than two requires costly sorting of the input primitives, and even produces wrong results in case of intersecting primitives.
Several techniques have been developed relying on even more costly per-pixel sorting to circumvent this problem, either by sorting explicitly~\cite{carpenter1984abuffer} or implicitly using multiple render passes~\cite{everitt2001interactive,bavoil2008dual}. These algorithms are in general invariant to the order of input primitives and produce correct results, and are therefore called \emph{exact order-independent transparency (OIT)} methods. A good overview can be found in~\cite{Wyman2016}.
A plethora of approaches try to approximate exact OIT by avoiding the costly sorting step. 
Some of them are not truly order-independent: the outcome still varies slightly depending on the order of input primitives~\cite{bavoil2007kbuffer,enderton2010stochastic,salvi2011adaptive,salvi2014multi}.
True, \emph{approximate OIT} algorithms generally modify the blending operations to achieve order-invariance.
Ignoring the order-dependent parts of the alpha blending formula results in the ``weighted sum'' operator~\cite{Meshkin2007}, which works well with surfaces of small opacity but becomes increasingly inaccurate for more opaque surfaces, causing over-darkening or over-brightening.
The weighted blended order-independent transparency (WBOIT) operator avoids the over-darkening by replacing the opacity and surface colors with opacity-weighted averages and introducing a monotonic decreasing depth weighting factor, heuristically reducing the influence of distant surfaces~\cite{McGuire2013Transparency}.
This approach works well for surfaces of low opacity and similar color, like smoke, but becomes problematic for almost opaque surfaces, and the weighting function needs to be tuned for every scene to achieve optimal results.
Extending this approach to rendering the surfaces into layers and applying the WBOIT operator to each before blending the resulting colors in order provides an interesting trade-off between quality and efficiency~\cite{Friederichs2021layered}.
Moment-based transparency techniques remedy some of the shortcomings of WBOIT by computing moments of the transmittance which are used in a second pass to reconstruct an approximate transmittance function, that weighs the influence of each surface on the final color~\cite{Muenstermann2018}.
Sharp changes in the transmittance function are not well captured at a lower number of power moments, whereas higher numbers decrease the computational efficiency.
As the influence of later surfaces is generally less than the first surfaces due to decreasing transmittance, another, truly order-independent approach is \textit{hybrid transparency} that blends the first $K$ surfaces correctly (including sorting), whereas the remaining surfaces are blended without sorting~\cite{maule2013hybrid}.
In this work, we integrate this hybrid transparency approach into the 3DGS framework and extend the formulation proposed by Maule~et~al.\ to improve its behavior within a gradient descent-based optimization scheme.

\section{Method}
We introduce the employed scene representation and then describe the two parts of our contribution.
\subsection{Preliminaries}
\label{ssec:3dgs_prelim}
Following Kerbl~et~al.~\cite{kerbl3Dgaussians}, we represent the scene using a set of 3D points with anisotropic Gaussian extent as a proxy for geometry.
Cutting each Gaussian at $3\sigma$ results in ellipsoid-shaped splats.
Each splat is defined by its 3D mean $\vect{\mu}\in\mathbb{R}^3$, a scalar opacity value $o\in\mathbb{R}$, three principal tangential vectors $\vect{t}_u, \vect{t}_v, \vect{t}_w\in\mathbb{R}^3$ modeling its orientation, and three scalars $s_u, s_v, s_w\in\mathbb{R}$ modeling its scale.
Note that $\vect{t}_u, \vect{t}_v, \vect{t}_w$ represent a 3D rotation, which allows modeling these parameters using a quaternion $q\in\mathbb{R}^4$ that is easier to optimize via gradient descent.
For rendering, we obtain a valid opacity value in $[0, 1]$ using a Sigmoid activation, represent $s_u, s_v, s_w$ in logarithmic space, and ensure $q$ is normalized.
The view-dependent color representation based on spherical harmonics (SH) is identical to 3DGS, i.e., each point uses SH up to degree 3 resulting in 48 coefficients total.

For any point $\vect{x}\in\mathbb{R}^3$, we can obtain an $\alpha$ value for a given 3D Gaussian by multiplying its opacity with the value of the Gaussian at that point:
\begin{equation}\label{eq:gauss_alpha_computation}
\alpha = o \cdot \text{e}^{-\frac{1}{2}\rho(\vect{x})^2} \quad \text{using} \quad \rho(\vect{x})^2 = (\vect{x} - \vect{\mu})^T\Sigma^{-1}(\vect{x} - \vect{\mu}),
\end{equation}
where $\Sigma^{-1}$ can easily be computed from $(\vect{t}_u, \vect{t}_v, \vect{t}_w)$ and $(s_u, s_v, s_w)$ (cf.~Kerbl et al.~\cite{kerbl3Dgaussians}).
These $\alpha$ values are then used in conjunction with the view-dependent RGB color $\vect{c}\in\mathbb{R}^3$ when computing per-pixel colors through standard depth-sorted alpha blending of all $N$ contributing splats:
\begin{equation}\label{eq:alpha_blending}
C = \sum_{i=1}^{N} \alpha_i  T_i \vect{c}_i, \quad \text{with} \quad T_i = \prod_{j=1}^{i-1} (1 - \alpha_j).\ 
\end{equation}

\begin{figure*}[thb]
\centering
\setlength\tabcolsep{0pt}
\begin{tabular}{@{}lr}
 \multirow{5}{*}{\includegraphics[width=.9\linewidth]{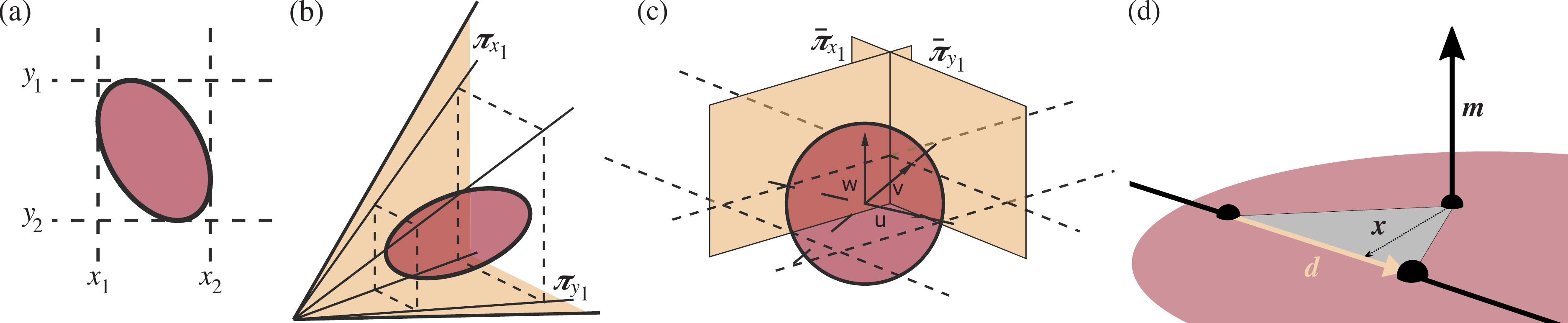} }&\vspace{15pt}\\
   & $\|\vect{m}\| = 2A$\vspace{15pt}\\
   & $A = \frac{\|\vect{d}\|\|\vect{x}\|}{2}$\vspace{15pt}\\
   &$\rho=\|\vect{x}\|$\\
   &\\
\end{tabular}
\caption{%
The perspectively correct screen-space bounding box of a splat (a) is given by the projection of its bounding frustum in view space (b). When transformed into local splat coordinates, the frustum planes align with tangential planes of the unit sphere (c). Our approach for splat evaluation along viewing rays makes use of the Pl{\"u}cker coordinate representation $( \vect{d} : \vect{m} )$. In local splat coordinates, the point along the ray that maximizes the Gaussian's value corresponds to the point $\vect{x}$ that minimizes the perpendicular distance $\|\vect{x}\|$ to the origin (d).
Parts (a-c) courtesy of Weyrich~et~al.~\cite{weyrich2007hardware}; used with permission.
}\label{fig:Pluecker}
\end{figure*}

\subsection{Accurate Splat Bounding and Evaluation}
\label{ssec:splat_eval}
In 3DGS, Kerbl~et~al.~\cite{kerbl3Dgaussians} project a 3D Gaussian onto the image plane using the Jacobian of the affine approximation of the projective transformation \cite{zwicker2001ewasplat}.
The result is a 2D Gaussian that can easily be evaluated for each pixel and also allows for easy construction of an axis-aligned bounding box around the center (e.g., at $3\sigma$) in screen space for accelerated computation.
However, this approach is not fully perspective-accurate, meaning it introduces artifacts when splats are viewed from certain angles (see \cref{fig:persp_correct}).

In this work, we propose a solution for this problem that builds upon an established approach for perspectively accurate rendering of ellipsoidal surfaces by Sigg~et~al.~\cite{sigg2006quadrics} and the respective optimizations by Weyrich~et~al.~\cite{weyrich2007hardware} for 2D~Gaussian splats.
It should be noted that these approaches elegantly avoid the matrix inversions required by similar approaches (e.g., \cref{eq:gauss_alpha_computation}), which makes them numerically stable even for degenerate splats, i.e., those where one or more main axes vanish.
Notably, Huang~et~al.~\cite{huang20242d} also base their perspectively accurate ray-splat intersection and evaluation on the aforementioned work.
However, their approach only works for 2D~Gaussians, i.e., degenerate splats.
In contrast, we now introduce an approach that enables perspectively accurate 3D splat evaluation along per-pixel viewing rays.
Thus, our approach is more general compared to Huang~et~al., as the 3D~Gaussians we use may still degenerate to 2D~Gaussians without impacting the optimization or rendering due to the numerical stability of our approach.

We start by describing the approach for computing tight, axis-aligned screen-space bounding boxes for each 3D Gaussian.
We define the model-view matrix $M\in\mathbb{R}^{4\times4}$, the projection to clip-space $P\in\mathbb{R}^{4\times4}$, and the viewport transform $V\in\mathbb{R}^{4\times4}$ mapping to window coordinates.
Using the per-splat attributes described in \cref{ssec:3dgs_prelim}, the transformation $T'\in\mathbb{R}^{4\times4}$ from the normalized splat space (where the Gaussian's ellipsoid becomes the unit sphere) to screen space is:
\begin{equation}\label{eq:T_matrix}
T' = VPMT, \quad \textrm{where} \quad T = \begin{pmatrix}
\topVert & \topVert & \topVert & \topVert \\
s_u \vect{t}_u & s_v \vect{t}_v & s_w \vect{t}_w & \vect{\mu} \\
\botVert & \botVert & \botVert & \botVert \\
 0 & 0 & 0 & 1
\end{pmatrix}
\end{equation}
is the transformation from normalized splat space to world space.
Using Weyrich~et~al.'s optimization of Sigg~et~al.'s bounding box computation~\cite{sigg2006quadrics}, extended to non-degenerate (full~3D) Gaussians, the desired tight 3D bounding box in screen space is $[b_1, t_1]\times[b_2, t_2]\times[b_3, t_3]$ (bottom and top values, $b_i$ and $t_i$, for $i\in\{1,2,3\}$ the $x$, $y$, and $z$ coordinate, respectively) is:
\begin{align}
b_i &= p_i - h_i
\\
t_i &= p_i + h_i
\;,
\end{align}
with
\begin{align}
s &= \langle (\rho_c,\rho_c,\rho_c,-1), T'_4 \odot T'_4 \rangle\;,\quad
\rho_c = 2\ln{\frac{o}{\tau_\alpha}}\;,\\
\vect{f} &= \frac{1}{s}(\rho_c,\rho_c,\rho_c,-1)\;,\\
p_i &= \langle\vect{f}, T'_i \odot T'_4\rangle\;,\\
h_i &= \sqrt{ p_i^2 - \langle\vect{f}, T'_i \odot T'_i\rangle }\;,\label{eq:hi}
\end{align}
and $T'_i$ the $i$-th row of $T'$, $\odot$ denoting component-wise multiplication, and the dot product denoted by $\langle\cdot,\cdot\rangle$.
A key advantage of this calculation, for which we provide a visual explanation in \cref{fig:Pluecker}, is that it offers a closed-form solution without the need for matrix inversion and regardless of whether the ellipsoid is degenerate in any direction.
Following Radl~et~al.~\cite{radl2024stopthepop}, we also compute an individual cutoff value for each splat to obtain smaller bounding boxes when their opacity value is less than one based on $\tau_\alpha$, a hyperparameter defining the minimum $\alpha$ value of a fragment for it not to be skipped during blending (see \cref{eq:gauss_alpha_computation}).

Next, we seek to compute the value of the 3D Gaussian for a given viewing ray.
In computer graphics, ray intersections with deformed primitives, such as our 3D splats, are an established practice since the dawn of ray tracing, generally following the approach of inversely transforming a viewing ray into the undeformed space.
In our case, re-using quantities of the bounding-box setup above, this would mean transforming the ray by $(VPMT)^{-1}$ into splat space, where the distance of the transformed ray to the origin yields $\rho$, \cref{eq:gauss_alpha_computation}.
In order to avoid the matrix inversion that we successfully sidestepped during bounding-box calculation, we once again follow Weyrich~et~al.\ and represent a viewing ray through a pixel at $(x_\text{s},y_\text{s})$ as two perpendicular planes $\vect{\pi}_x=(1,0,0,-x_\text{s})^{\!\top\!}$ and $\vect{\pi}_y=(0,1,0,-y_\text{s})^{\!\top\!}$ in screen space. Note that we follow the common convention of representing planes as homogeneous vectors $\vect{p}=(a, b, c, d)^{\!\top}$, so that a point $\vect{x}$ lies on the plane if and only if $\vect{p}^{\!\top}\vect{x}=0$.
In contrast to 3D points, transforming those planes by $(VPMT)^{-1}$ into planes $\vect{\pi}^\text{s}_{x/y}$ in splat space requires the inverse-transposed mapping, that is:
\begin{align}
\label{eq:ray-mapping}
\vect{\pi}^\text{s}_{x/y} &= \bigl((VPMT)^{-1}\bigr)^{-\!\top}\vect{\pi}_{x/y} = (VPMT)^{\!\top}\vect{\pi}_{x/y}
\;,
\end{align}
once again sidestepping inversion.
Next up, intersection of $\vect{\pi}^\text{s}_x$ and $\vect{\pi}^\text{s}_y$ yields the viewing ray in splat space, and its distance to the origin is $\rho$.
In contrast to Weyrich~et~al., however, $\rho$ cannot quite as easily be computed, due to the non-degeneracy of our 3D Gaussians.
Instead, we employ the dual Pl{\"u}cker coordinate representation that derives the mapped viewing ray $\mathcal{L}^*_\text{s}$ from the intersection of $\vect{\pi}^\text{s}_{x/y}$ from their coefficients $\vect{\pi}^\text{s}_x =: (a^1 : a^2 : a^3 : a^0)$, and $\vect{\pi}^\text{s}_y =: (b^1 : b^2 : b^3 : b^0)$, respectively:
\begin{align}
\mathcal{L}^*_\text{s} &= (  p^{23} : p^{31} : p^{12} : p^{01} : p^{02} : p^{03} )
\;,\\
\text{with}\quad
p^{ij} &= \begin{vmatrix}a^i & a^j\\ b^i & b^j\end{vmatrix} = a^i b^j - a^j b^i.
\end{align}
The line's dual Pl{\"u}cker coordinates $\mathcal{L}^*_\text{s}$ are numerically equivalent to its primal Pl{\"u}cker coordinates $\mathcal{L}_\text{s}$ up to some common scaling factor $\lambda$:
\begin{align}
\mathcal{L}_\text{s} =: ( \vect{d} : \vect{m} ) &= ( p_{01} : p_{02} : p_{03} : p_{23} : p_{31} : p_{12} )
\;\\
&=( \lambda p^{23} : \lambda p^{31} : \lambda p^{12} : \lambda p^{01} : \lambda p^{02} : \lambda p^{03} )
\;
\end{align}
where $\vect{d}$ is the ray direction in splat space and $\vect{m}$ is its moment around the origin (see \cref{fig:Pluecker} for a visualization).
From the known equality $\textit{distance}(\mathcal{L}_\text{s}, \textit{origin}) = \frac{\|\vect{m}\|}{\|\vect{d}\|}$, it follows that:
\begin{align}
\label{eq:rho-pluecker}
\rho(\vect{x})^2 &= 
\frac{\| \lambda (p^{01}, p^{02}, p^{03})^{\!\top} \|^2}{\|\lambda (p^{23}, p^{31}, p^{12})^{\!\top}\|^2}
=
\frac{
\begin{Vmatrix}
a^0b^1 - a^1b^0\\
a^0b^2 - a^2b^0\\
a^0b^3 - a^3b^0
\end{Vmatrix}^2
}
{
\begin{Vmatrix}
a^2b^3 - a^3b^2\\
a^3b^1 - a^1b^3\\
a^1b^2 - a^2b^1
\end{Vmatrix}^2
}
\;.
\end{align}
Note that we directly compute $\rho(\vect{x})^2$ for \cref{eq:gauss_alpha_computation}.
This formulation is numerically stable, as the only potential issue -- a vanishing denominator in~\cref{eq:rho-pluecker} -- can easily be detected and corresponds to the ray missing the splat. 
Conveniently, the obtained value corresponds to the point along the pixel's viewing ray where the Gaussian has the highest value, essentially making it equivalent to the numerically unstable and slower approaches used in concurrent works~\cite{yu2024gof, moenne2024gaussiantracer}.
Compared to Kerbl~et~al.~\cite{kerbl3Dgaussians}, our calculations incur reduced setup costs for each Gaussian but slightly higher costs per pixel, which is in fact a desirable tradeoff whenever graphics primitives cover only few pixels~\cite{Montrym97InfiniteReality}.
Moreover, the added computations are exclusively additions and multiplications, except for the division in~\cref{eq:rho-pluecker}, meaning they are both fast and trivial to differentiate.

\subsection{Temporally-Stable Rendering via Hybrid Transparency}
\label{ssec:hybrid_transp}
In 3DGS, Kerbl~et~al.~\cite{kerbl3Dgaussians} use a tile-based rasterization approach to efficiently render images from a set of 3D Gaussian primitives.
To achieve robust optimization via gradient descent, they compute per-pixel color by applying standard $\alpha$-blending,~\cref{eq:alpha_blending}, in depth-sorted order.
In this work, we propose to integrate a hybrid-transparency approach for blending into the 3DGS framework.
This is motivated by three observations:
(1) The global sorting scheme proposed by Kerbl~et~al.~\cite{kerbl3Dgaussians} sorts splats as a whole, leading to incorrectly resolved splat-splat intersections and temporal incoherence (``popping'') during rendering and optimization; resolving these artifacts through a full sort of all 3D Gaussian pixel fragments, however, is prohibitively slow and thus approximate sorting is required (cf.~Radl~et~al.~\cite{radl2024stopthepop}).
(2) As we will demonstrate in our experiments, using ``full'' OIT, which forgoes sorting completely, works well during optimization and even improves background reconstruction but results in semi-transparent foreground rendering, which is clearly undesirable for novel-view synthesis and reconstruction methods.
(3) Blending only the first $K$ contributions for each pixel has been shown to work well in differentiable rendering frameworks while also much faster than approaches that require a complete sort~\cite{lassner2021pulsar, franke2024trips}.

The hybrid-transparency approach~\cite{maule2013hybrid} splits up the per-pixel color computation into two parts:
The \textit{core} that uses standard $\alpha$-blending to blend the first $K$ contributions for each pixel, including sorting.
All remaining contributions are combined into a \textit{tail}, that is blended without sorting.
To decide whether a splat is one of the first $K$, we need to compute a pixel-specific depth value for each of them, i.e., the $z$-value of the point of evaluation $\vect{x}$ (see \cref{fig:Pluecker}) in view space.
Re-using the intermediate values of our splat evaluation (see \cref{ssec:splat_eval}), $\vect{x}_\text{view}$ can be computed efficiently by
\begin{equation}
\vect{x}_\text{view} = MT\frac{\vect{d}\times\vect{m}}{\|\vect{d}\|^2}.
\end{equation}
As in \cref{ssec:splat_eval}, $\vect{x}_\text{view}$ is the splat's point of maximum contribution along the pixel ray, which again makes our computation compatible with the numerically unstable and slower approaches used in recent work~\cite{radl2024stopthepop}.
In our implementation, we keep track of the first $K$ contributions and accumulate all of the ($N-K$) contributions for each pixel to then compute its color as
\begin{equation}\label{eq:pixel_color_computation}
C = \sum_{i=1}^{K} \alpha_i  T_i \vect{c}_i \; +\;  T_{K+1}\Bigl( (1 - T_\text{tail}) \vect{c}_\text{tail} \; + \; T_\text{tail} \vect{c}_\text{bg} \Bigr)\;,
\end{equation}
with
\begin{equation}\label{eq:pixel_color_computation_helper}
T_i = \prod_{j=1}^{i-1} (1 - \alpha_j),\;\;\;
\vect{c}_\text{tail} = \frac{\sum_{i=K+1}^N \alpha_i \vect{c}_i }{\sum_{i=K+1}^N \alpha_i},\;\;\;
T_\text{tail} = \prod_{\mathclap{i=K+1}}^N (1 - \alpha_i)\;.
\end{equation}
Notably, our computation of $T_\text{tail}$ is more accurate compared to the average-based formulation of Maule~et~al.~\cite{maule2013hybrid}.
As the partial derivative of the blending function is the same for all splats inside the tail, our approach allows pre-computing all required partial derivatives for each pixel during the forward pass.
Thus, our backward pass can be implemented very efficiently.

\section{Experiments}
We conduct multiple experiments on established datasets to validate our approach.

\begin{table*}[tb]
\centering
\setlength\tabcolsep{5.4pt}
\small
\begin{tabular}{@{}lcccccc|ccccccccc}
\toprule
                                           & \multicolumn{6}{c|}{Mip-NeRF360~\cite{barron2022mipnerf360} (Img. Res. $\approx$ 1MP--1.5MP)}                                            & \multicolumn{6}{c}{Tanks\&Temples~\cite{Knapitsch2017} (Img. Res. $\approx$ 2MP)}                                                        \\
Method                                     & SSIM$\uparrow$       & PSNR$\uparrow$       & LPIPS$\downarrow$    & Training              & Render               & \#Splats             & SSIM$\uparrow$       & PSNR$\uparrow$       & LPIPS$\downarrow$    & Training              & Render               & \#Splats             \\ 
\midrule
Zip-NeRF~\cite{barron2023ICCV}             & 0.828 & 28.56 & 0.219 & 5h & 5s & N$\slash$a & 0.878 & 26.75 & 0.233 & 5h & 10s & N$\slash$a \bstrut\\
\hline
3DGS~\cite{kerbl3Dgaussians}               & \cellcolor{3rd}0.814 & 27.20 & 0.254 & \cellcolor{3rd}18m40s & \cellcolor{3rd}3.2ms  & 3.31M & \cellcolor{1st}0.866 & \cellcolor{1st}25.27 & 0.276 & \cellcolor{3rd}20m18s & \cellcolor{3rd}4.6ms  & 1.50M \tstrut\\
StopThePop~\cite{radl2024stopthepop}       & \cellcolor{3rd}0.814 & \cellcolor{2nd}27.30 & \cellcolor{3rd}0.252 & 20m11s & 4.7ms  & 3.28M & \cellcolor{1st}0.866 & \cellcolor{3rd}24.99 & \cellcolor{1st}0.270 & 22m03s & 4.8ms  & 1.49M \\
~$\hookrightarrow$ w$\slash$ opacity decay & 0.809 & 27.08 & 0.267 & \cellcolor{2nd}12m02s & \cellcolor{2nd}2.5ms & \cellcolor{1st}1.71M & \cellcolor{2nd}0.862 & 24.98 & 0.284 & \cellcolor{2nd}14m22s & \cellcolor{2nd}3.0ms & \cellcolor{1st}0.81M \\
2DGS~\cite{huang20242d}                  & 0.795 & 26.81 & 0.297 & 21m08s & 7.4ms  & \cellcolor{2nd}2.01M & \cellcolor{3rd}0.851 & 24.55 & 0.314 & 28m45s & 10.4ms & \cellcolor{2nd}0.95M \\
Huang~et~al.~\cite{huang2024erroranalysis} & \cellcolor{3rd}0.814 & \cellcolor{1st}27.41 & 0.257 & 1h01m  & 10.7ms & 3.30M & \cellcolor{1st}0.866 & \cellcolor{2nd}25.19 & 0.276 & 1h13m  & 15.7ms & 1.49M \\
GOF~\cite{yu2024gof}                       & \cellcolor{2nd}0.821 & \cellcolor{3rd}27.27 & \cellcolor{2nd}0.238 & 1h21m  & 60.0ms & 2.91M & \cellcolor{1st}0.866 & 25.01 & \cellcolor{3rd}0.274 & 1h35m  & 81.3ms & \cellcolor{3rd}1.26M \\
Ours                                       & \cellcolor{1st}0.822 & 27.17 & \cellcolor{1st}0.234 & \cellcolor{1st}10m07s & \cellcolor{1st}2.2ms & \cellcolor{3rd}2.19M & \cellcolor{1st}0.866 & 24.62 & \cellcolor{2nd}0.272 & \cellcolor{1st}8m33s  & \cellcolor{1st}2.6ms & \cellcolor{1st}0.81M \\
\bottomrule
\end{tabular}
\caption{%
Quantitative comparisons on the Mip-NeRF360 and Tanks and Temples datasets. Our approach of using perspectively correct splat evaluation in combination with hybrid transparency significantly reduces training and rendering times with image quality being similar to the baselines'. Excluding Zip-NeRF, the three best results are highlighted in \textcolor{1stText}{\textbf{green}} in descending order of saturation.
}\label{tab:main_results}
\end{table*}

\subsection{Setup}
\label{ssec:setup}
Following recent work, we evaluate on the established Mip-NeRF360~\cite{barron2022mipnerf360} and Tanks and Temples~\cite{Knapitsch2017} datasets, which provide a diverse set of 17 real scenes with varying challenges.
We use all nine scenes from the Mip-NeRF360 dataset as well as all eight scenes from the \textit{intermediate} set of the Tanks and Temples dataset resulting in a total of 17 scenes.
This allows us to evaluate our method on a broad spectrum of challenges regarding both geometric and photometric aspects.
As is common practice, we use the $4\times \slash \ 2\times$ downscaled images for the outdoor$\slash$indoor scenes of the Mip-NeRF360 dataset.
For the Tanks and Temples scenes, we use the full-resolution images to evaluate performance at full HD resolution.
To compute meaningful quality metrics, we use the established 7:1 train$\slash$test split~\cite{barron2022mipnerf360} for all scenes.
Central to our evaluation is the comparison against the original 3D Gaussian Splatting~\cite{kerbl3Dgaussians}.
Furthermore, we compare against the recent StopThePop~\cite{radl2024stopthepop}, 2D Gaussian Splatting~\cite{huang20242d}, and Huang~et~al.'s approach~\cite{huang2024erroranalysis}, as they also provide solutions for the problems addressed in this work.
For StopThePop, we report results of both model variants the authors propose in their paper~\cite{radl2024stopthepop}.
We also compare against Gaussian Opacity Fields (GOF)~\cite{yu2024gof}, a very recent work that uses ray tracing-based volume rendering.
For all baselines, we use the official implementation to compute the results for all 17 scenes and use the same script for all quality metric computations.
We use a single RTX 4090 for all experiments and measure frame rate by rendering the test-set images of each scene 100 times and averaging over the results~\cite{duckworth2023smerf}.

\subsection{Implementation and Optimization}
\label{ssec:implementation}
We implement our method from scratch in PyTorch and CUDA using the implementation of Kerbl~et~al.~\cite{kerbl3Dgaussians} as a reference.
While the optimization pipeline and hyperparameters are mostly the same as in 3DGS, we make multiple modifications to solve technical challenges arising from our changes to the rendering pipeline.
First, we tackle aliasing problems by employing our own fast CUDA implementation of the 3D filter proposed by Yu~et~al.~\cite{yu2024mip} as our inversion-free splat evaluation prevents use of the screen-space low-pass filter in EWA splatting~\cite{zwicker2001ewasplat}.
As the screen-space gradients 3DGS uses for densification are not available with our splat evaluation scheme, we follow Moenne-Loccoz~et~al.\ and instead directly use the gradient of the splat means scaled by half the respective distance to the camera in view space~\cite{moenne2024gaussiantracer}.
To prevent over-densification, we replace the opacity reset with a decay strategy~\cite{radl2024stopthepop} that multiplies splat opacity by $\lambda_o=0.9995$ every $50$ iterations during densification.
Furthermore, we adapt the densification changes from GOF to our approach~\cite{yu2024gof} and apply visibility score-based pruning as proposed by Niemeyer~et~al.~\cite{niemeyer2024radsplat}.
As we will show in our experiments, these changes are necessary to adjust 3DGS’s densification strategy to work with our ray-splat intersection-based approach.
For our blending formulation, we use a core with $K=16$ and employ a second threshold $\tau_K=0.05$ specifying the minimum $\alpha$ of a fragment to be considered for the core.
As the insertion sort that we employ for keeping track of per-pixel core causes significant divergence, we reduce the tile size to $8\times8$ in our implementation.
Lastly, we make use of an optimized SSIM implementation~\cite{taming3dgs} to speed up loss computation.

\subsection{Results}
\label{ssec:results}
The quantitative results in \cref{tab:main_results} show that our approach provides significantly faster training and rendering.
Compared to 3DGS~\cite{kerbl3Dgaussians}, we achieve a training speedup of $2.1\times$ and render $1.6\times$ faster on average.
The second fastest approach is StopThePop~\cite{radl2024stopthepop} with opacity decay enabled, which benefits from a large reduction in the number of splats.
In terms of image quality, we observe that our method achieves about equal results across the board.
However, all 3DGS-based approaches are clearly outperformed by Zip-NeRF~\cite{barron2023ICCV}.
Taking our contributions into consideration, we observe that our approach outperforms approaches with perspectively correct splat evaluation.
Compared to 2DGS~\cite{huang20242d}, we achieve better results across the board reassuring that 3D Gaussian splats are more expressive than the 2D Gaussian surfels used by 2DGS.
Huang~et~al.'s projection method~\cite{huang2024erroranalysis} significantly slows down training and rendering while achieving image quality similar to that of 3DGS.
GOF~\cite{yu2024gof} achieves slightly higher image quality but requires longer training and fails to render in real-time.
It should be noted that 2DGS and GOF were fine-tuned for mesh extraction after optimization, which is an interesting topic but not a focus of this work.
Seeing that we achieve similar image quality but faster training and rendering compared to either configuration of StopThePop~\cite{radl2024stopthepop}, the results confirm that our approach is a valid alternative to prevent popping artifacts for rasterization-based 3D Gaussian splat rendering.

\begin{figure*}[tbh]
\centering
{\footnotesize
\setlength\tabcolsep{0pt}
\begin{tabular}{*{6}{p{0.166666666667\linewidth}<{\centering}}@{}}
3DGS~\cite{kerbl3Dgaussians} & Huang~et~al.~\cite{huang2024erroranalysis} & StopThePop~\cite{radl2024stopthepop} & 2DGS~\cite{huang20242d} & Ours & Ground Truth
\end{tabular}
}
\includegraphics[width=\linewidth]{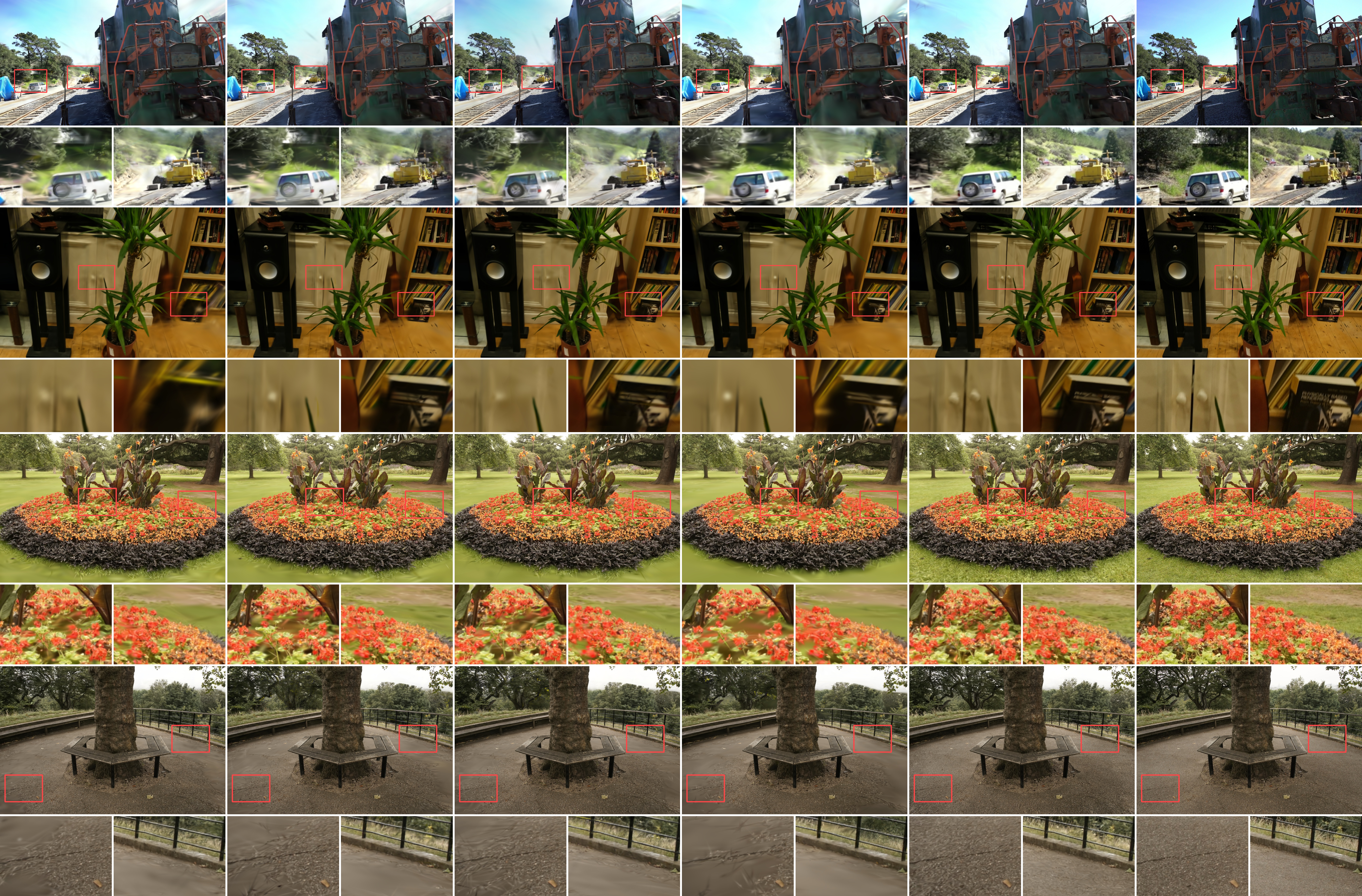}
\caption{%
Visual comparisons of baselines that allow for real-time rendering.
}\label{fig:vis_comp}
\end{figure*}

We also show visual comparisons for multiple scenes in \cref{fig:vis_comp}.
Beyond those comparisons, we often observe more accurate reconstruction of the background when inspecting trained models.
This is likely caused by our perspectively correct splat evaluation that allows better use of the multi-view signal provided by the training images where the background is generally located near image borders where perspective distortion has its highest influence.

\noindent\textbf{Performance Breakdown.}
In \cref{tab:render_timings}, we compare how parts of the rendering influence the total runtime.
We denote the creation of per-tile primitive lists (cf. Kerbl~et~al.~\cite{kerbl3Dgaussians}) as \textit{tiling} and per-tile color computations as \textit{blending}.
The speed of 3DGS is equally limited by tiling and blending.
StopThePop~\cite{radl2024stopthepop} implements significant optimizations for the tiling step in an attempt to reduce the overhead of their hierarchical sorting during blending, which is the main bottleneck.
In contrast, we use a 16-bit unsigned integer as the key for each Gaussian/tile instance, as we do not require any global depth-sorting.
This speeds up tiling by a similar amount as the revised culling strategy of StopThePop.
Regarding the blending step, our hybrid transparency approach requires only partial depth-ordering making it significantly faster than the approach of StopThePop.

\begin{table}[tbh]
\centering
\setlength\tabcolsep{3.8pt}
\small
\begin{tabular}{@{}lccccc}
\toprule
Method                                     & Preprocess     & Tiling         & Blending       & Total          & \#Splats       \\
\midrule
                                           & \multicolumn{5}{c}{Bicycle~\cite{barron2022mipnerf360}}                            \\ \cmidrule{2-6}
3DGS~\cite{kerbl3Dgaussians}               & 1.099          & 3.494          & 3.422          & 8.015          & 6.03M          \\
StopThePop~\cite{radl2024stopthepop}       & 1.364          & 1.358          & 5.116          & 7.838          & 6.05M          \\
~$\hookrightarrow$ w$\slash$ opacity decay & \textbf{0.554} & \textbf{0.547} & 2.442          & 3.543          & \textbf{2.91M} \\
2DGS~\cite{huang20242d}                  & 0.674          & 1.909          & 5.267          & 7.850          & 3.99M          \\
Ours                                       & 0.556          & 0.754          & \textbf{2.007} & \textbf{3.317} & 4.35M          \\
\midrule
                                           & \multicolumn{5}{c}{Francis~\cite{Knapitsch2017}}                                   \\ \cmidrule{2-6}
3DGS~\cite{kerbl3Dgaussians}               & 0.235          & 2.078          & 1.644          & 3.957          & 0.81M          \\
StopThePop~\cite{radl2024stopthepop}       & 0.268          & 0.423          & 2.869          & 3.560          & 0.82M          \\
~$\hookrightarrow$ w$\slash$ opacity decay & 0.154          & \textbf{0.310} & 1.980          & 2.444          & 0.45M          \\
2DGS~\cite{huang20242d}                  & 0.177          & 4.397          & 8.700          & 13.27          & 0.51M          \\
Ours                                       & \textbf{0.107} & 0.678          & \textbf{1.228} & \textbf{2.013} & \textbf{0.44M} \\
\bottomrule
\end{tabular}
\caption{%
Breakdown of render timings in milliseconds at a resolution of 1920{\texttimes}1080 pixels measured on an RTX 4090 GPU. We select the scenes where the trained 3DGS model contains the highest (Bicycle) and lowest (Francis) number of splats across all tested scenes. For each method, we repeatedly render all test-set views and average measurements obtained via NVIDIA Nsight Systems across 100 runs.
}\label{tab:render_timings}
\end{table}

\noindent\textbf{Perspectively Correct Splat Evaluation.}
As previously discussed, our contributions can be integrated into 3D Gaussian splatting implementations independently.
Our hybrid transparency approach (\cref{ssec:hybrid_transp}) is mainly a method for addressing popping artifacts without sacrificing computational efficiency.
However, our perspectively accurate splat evaluation (\cref{ssec:splat_eval}) should prove beneficial across a much broader range of applications due to its accuracy.
As can be seen in \cref{tab:splat_eval}, using it alongside the approximately depth-ordered alpha blending from 3DGS~\cite{kerbl3Dgaussians} improves results for all image quality metrics. 
Note that a major portion of the speed increase over 3DGS comes from our approach providing tight screen-space bounds for each splat. These significantly reduce the amount of global memory accesses, sorting overhead, and the number of splats processed during blending.

\begin{table}[htb]
\centering
\setlength\tabcolsep{4.0pt}
\small
\begin{tabular}{@{}lccccc}
\toprule
Method                & SSIM$\uparrow$ & PSNR$\uparrow$ & LPIPS$\downarrow$ & Training        & Render          \\ \midrule
3DGS                  & 0.730          & 24.64          & 0.265             & 21m17s          & 3.41ms          \\
Ours                  & 0.742          & 24.62          & 0.237             & \textbf{12m09s} & \textbf{2.08ms} \\
Ours w/o per-pixel HT & \textbf{0.748} & \textbf{24.81} & \textbf{0.230}    & 17m26s          & 2.23ms          \\ 
\bottomrule
\end{tabular}
\caption{%
We showcase the potential of our perspectively correct splat evaluation by omitting our hybrid transparency-based blending in favor of the approximately depth-ordered alpha blending from 3DGS~\cite{kerbl3Dgaussians}. As it forgoes per-pixel ordering, this version of our model is subject to popping artifacts, and it is slightly less efficient than our default configuration, but outperforms 3DGS across all metrics. The reported values are averaged results for the five outdoor scenes from the Mip-NeRF360 dataset~\cite{barron2022mipnerf360}.
}\label{tab:splat_eval}
\end{table}

\noindent\textbf{Hybrid Transparency Ablations.}
To analyze the behavior of our hybrid transparency-based blending, we perform ablations on the nine scenes from the Mip-NeRF360 dataset~\cite{barron2022mipnerf360}.
The quantitative results shown in \cref{tab:ablations} clearly indicate that $K=16$ is the optimal choice for our approach.
Note that rendering is slightly slower with $K=8$ as the number of splats created during optimization increases as core size decreases.

We also show visual comparisons in \cref{fig:ht_ablations} showing that blending without any sorting is not a valid option as it causes the foreground to become transparent.
However, we do find that background reconstruction seems to improve in this case.
With our default configuration of $K=16$, omitting tail computation during inference causes a small drop in image quality, which is most noticeable in the sky.
Importantly, not using the tail during training causes catastrophic failure.

\begin{table}[tbh]
\centering
\setlength\tabcolsep{2.6pt}
\small
\begin{tabular}{@{}lcccccc}
\toprule
Configuration              & SSIM$\uparrow$ & PSNR$\uparrow$ & LPIPS$\downarrow$ & Training          & Render          & \#Splats       \\
\midrule
                           & \multicolumn{6}{c}{Hybrid Transparency}                                                                    \\ \cmidrule{2-7}
$K=8$                      & 0.811          & 26.70          & 0.245             & \textbf{10m02s}   & 2.28ms          & 2.45M          \\
$K=32$                     & 0.821          & 27.12          & 0.236             & 31m11s            & 2.67ms          & \textbf{2.10M} \\
Ours ($K=16$)              & \textbf{0.822} & \textbf{27.17} & \textbf{0.234}    & 10m07s            & 2.19ms          & 2.19M          \\
Inference w$\slash$o tail  & 0.817          & 26.92          & 0.240             & 10m07s            & \textbf{2.14ms} & 2.19M          \\
\midrule
                           & \multicolumn{6}{c}{Densification}                                                                          \\ \cmidrule{2-7}
No 3D filter$^\dagger$     & 0.777          & 25.58          & 0.303             & \textbf{1m44s}    & 2.24ms          & 3.65M          \\
No opacity decay$^\dagger$ & 0.784          & 25.92          & 0.292             & 2m42s             & 3.81ms          & 4.35M          \\
No scaled grads            & 0.784          & 26.30          & 0.317             & 4m05s             & \textbf{1.35ms} & \textbf{0.30M} \\
No vis. pruning            & 0.820          & 26.96          & 0.235             & 12m17s            & 2.34ms          & 2.42M          \\
Ours-7K                    & 0.780          & 25.72          & 0.299             & 1m45s             & 2.10ms          & 2.09M          \\
Ours                       & \textbf{0.822} & \textbf{27.17} & \textbf{0.234}    & 10m07s            & 2.19ms          & 2.19M          \\
\bottomrule
\end{tabular}
\caption{%
Ablations for our hybrid transparency approach and our adaptation of the densification strategy from 3DGS~\cite{kerbl3Dgaussians} computed on the nine scenes from the Mip-NeRF360 dataset~\cite{barron2022mipnerf360}. Configurations marked with $\dagger$ are only trained for $7,000$ iterations as the optimization would otherwise run out of memory due to over-densification.
}\label{tab:ablations}
\end{table}

\noindent\textbf{Densification Ablations.}
As described in \cref{ssec:implementation}, we make multiple adjustments to the adaptive density control mechanism proposed by Kerbl~et~al.~\cite{kerbl3Dgaussians}.
The ablations in \cref{tab:ablations} clearly show why this is necessary for our ray-splat intersection-based approach.
Not scaling the gradient of the splat means by half the respective distance to the camera~\cite{moenne2024gaussiantracer} results in significant under-reconstruction, whereas dropping either the 3D filter~\cite{yu2024mip} or using the standard opacity reset~\cite{kerbl3Dgaussians} every 3,000 iterations instead of an opacity decay~\cite{radl2024stopthepop} results in the optimization running out of memory due to over-reconstruction.
Furthermore, the visibility score-based pruning~\cite{niemeyer2024radsplat} simultaneously improves quality and speed as it helps to prevents over-reconstruction by pruning low-opacity splats, which would otherwise cause overfitting.

\section{Discussion}
The experiments validate that our approach improves robustness and accuracy of the optimization due to our perspective-accurate and numerically stable splat evaluation.
Similarly, they show that hybrid transparency can effectively be integrated into differentiable rasterizers for 3D Gaussian splats to accelerate training and rendering without sacrificing image quality.
Importantly, both of our contributions improve the rendering quality beyond what shows in the quantitative results for standardized benchmarks.
To address this, our supplemental includes two videos that clearly demonstrate the advantage of our two-part contribution, which we encourage the reader to view.
The perspectively correct splat evaluation allows our rendering to avoid artifacts that arise from the approximate projection used in 3DGS~\cite{kerbl3Dgaussians}.
We find that the commonly used benchmark datasets largely fail to reveal this artifact making it invisible in quantitative evaluation.
However, when inspecting trained 3D Gaussian splat models in a real-time viewer, it is very easy to find scenarios where the perspectively inaccurate projection causes disturbing artifacts.

\begin{figure*}[thb]
\centering
\setlength\tabcolsep{0pt}
\begin{tabular}{*{6}{p{0.16666667\linewidth}<{\centering}}}
$K=0$ & $K=8$ & Ours & Ours & Render w$\slash$o Tail & Train w$\slash$o Tail
\end{tabular}
\includegraphics[width=\linewidth]{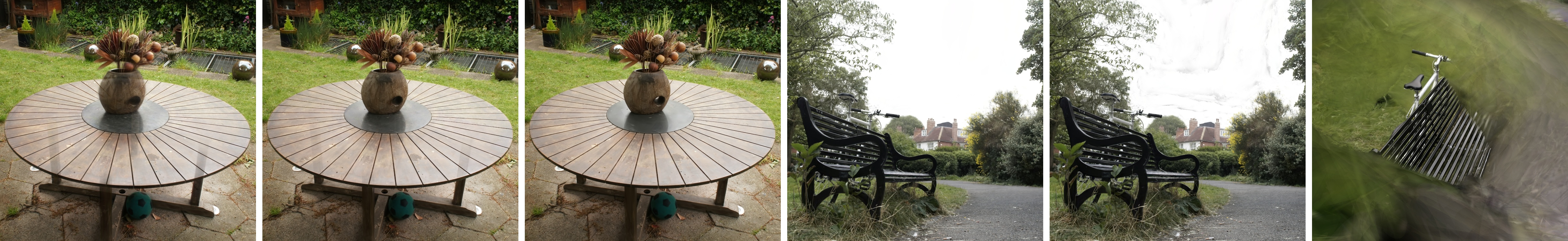}
\caption{%
Visual comparisons for different model configurations regarding our hybrid transparency approach. Using a smaller core size $K$ causes issues for reflective surfaces, as radiance fields commonly model these using semi-transparency. Disabling the order-independent tail only slightly reduces quality, especially in the sky, whereas not using it during optimization results in catastrophic failure.
}\label{fig:ht_ablations}
\end{figure*}

Equally as noticeable are the popping artifacts caused by the approximate depth-ordering in 3DGS.
StopThePop~\cite{radl2024stopthepop} uses a hierarchical sorting approach to resolve this issue.
They sort splats globally based on their means~\cite{kerbl3Dgaussians} and then use a sliding window approach to sort the per-tile lists approximately.
In contrast, our approach guarantees correct sorting order for the first $K$ elements in each pixel and then uses fast blending without sorting for the rest.
Our results show that this strategy results in a good tradeoff between quality and performance:
The correct sorting of the front-most fragments leads to high rendering quality and temporal stability, with little impact on performance.
While the remaining elements from the tail are essential for robust optimization as demonstrated in \cref{fig:ht_ablations}, our results show that the much cheaper unsorted blending is sufficient to maintain high performance, very good rendering quality, and robust optimization.

The artifacts addressed by our approach are most apparent while the camera is moving (see our supplemental videos), which is an issue for VR$\slash$AR applications or fast-paced games where the user rarely keeps their head or the camera perfectly still.
As VR also requires high frame rates to mitigate cybersickness and view-consistent peripheral rendering to avoid breaking immersion~\cite{franke2024vrsplatting}, we believe that the combined advantages of our approach -- fast, perspective-correct rendering without popping artifacts -- make it a great fit for such applications.
Another area where we think our approach could prove beneficial are 3DGS implementations for low-end devices, which currently require expensive CPU sorting.
As our approach naturally supports depth peeling due to the fixed size of the core, it should be possible to implement a hardware-accelerated viewer for our approach that does not require a dedicated sorting routine.

Regarding reconstruction and image quality, we observe that the adaptive density control mechanism by Kerbl~et~al.~\cite{kerbl3Dgaussians} introduced for 3DGS has a significant influence while being very sensitive to changes.
As outlined in \cref{ssec:implementation}, approaches evaluating splats along per-pixel viewing rays do not naturally provide the information used by the densification approach of 3DGS, i.e., the gradient of the splat's position in view-space.
After experimenting with the solutions presented in recent works~\cite{kheradmand243dgsmcmc, huang20242d, moenne2024gaussiantracer}, we find scaling the gradients of the 3D positions by half the distance to the camera~\cite{moenne2024gaussiantracer} works best for our approach.
Nonetheless, we find that the direct connection between local splat attributes and the global densification scheme introduces issues regarding the robustness of all 3DGS-based approaches we investigated.
Additional challenges, such as the photometric variation in images from the Tanks and Temples dataset~\cite{Knapitsch2017} further amplify this issue.

\noindent\textbf{Limitations and Future Work.}
Naturally, our approach is not without limitations.
While sole use of alpha blending allows for early stopping based on a preset transmittance threshold~\cite{kerbl3Dgaussians}, this is not an option for hybrid transparency by default.
In our current implementation, we accumulate the RGB color and alpha of all per-pixel contributions that are not part of the core.
Intuitively, this is a problem when scenes become larger due to the increase in depth complexity.
However, we are confident that this issue can be resolved, e.g., by combining a depth prepass with a depth-weighting function as in WBOIT~\cite{McGuire2013Transparency}.
Additionally, we find that our approach achieves slightly lower PSNR than that of 3DGS, which we think is due to the absence of a screen-space aliasing filter or multi-sampling approach.
As suggested by Huang~et~al.~\cite{huang20242d}, we tried using the approach of Botsch~et~al.~\cite{botsch2005high}, but did not find it beneficial.
We think it is possible to extend our splat evaluation to efficiently handle lens distortion thus enabling training on images without applying lossy undistortion operations during data pre-processing.
Lastly, we would like to highlight that our hybrid transparency approach unveils an avenue for integrating otherwise too expensive ideas into real-time rendering of 3D Gaussians splats.
While correct volume rendering of all contributing splats is not computationally feasible, it should be possible for the first $K=16$ Gaussians.
Similarly, we think that the fixed size of the core in our hybrid transparency approach allows the use of more powerful appearance models~\cite{duckworth2023smerf} to improve image quality.

\section{Conclusion}
In this paper, we addressed two significant challenges prevalent in 3D Gaussian splat rendering.
First, we introduced a novel approach for fast and perspective-accurate splat evaluation, which eliminates artifacts that arise from the affine approximation used by current 3D Gaussian Splatting implementations.
Our approach elegantly avoids matrix inversion thus ensuring robust optimization and accurate rendering without numerical instabilities, even when splats degenerate.
Second, we proposed an approach for view-consistent rendering through hybrid transparency, which uses fast, approximate per-pixel depth sorting that guarantees correct order for the first $K$ contributions and thus prevents popping artifacts.
Importantly, these improvements have a largely positive impact on performance, as evidenced by our analysis of training time, rendering speed, image quality metrics, and visual fidelity.
We believe our two-part contribution can both independently and jointly improve accuracy and efficiency of 3D Gaussian splat rendering in the future.

\begin{center}
\vspace{5pt}
\url{https://fhahlbohm.github.io/htgs/}
\end{center}

\section*{Acknowledgments}
We thank Timon Scholz and Carlotta Harms for their help with comparisons and supplemental material.
This work was partially funded by the DFG -- projects “Real-Action VR” (ID 523421583) and “Increasing Realism of Omnidirectional Videos in Virtual Reality” (ID 491805996) -- as well as the L3S Research Center, Hanover, Germany. Linus Franke was supported by the 5G innovation program of the German Federal Ministry for Digital and Transport under the funding code 165GU103B.

\bibliographystyle{eg-alpha-doi} 
\bibliography{references}  

\appendix
\section{Frame Rate Benchmarking}
\label{app_sec:fps_benchmarking}
Following Duckworth~et~al.~\cite{duckworth2023smerf}, we compute frame rate by rendering the test-set images of each scene 100 times and averaging over the results.
While the five compared baselines -- 3DGS~\cite{kerbl3Dgaussians}, StopThePop~\cite{radl2024stopthepop}, 2DGS~\cite{huang20242d}, GOF~\cite{yu2024gof}, and the approach of Huang~et~al.~\cite{huang2024erroranalysis} -- are all implemented on top of the original 3D Gaussian Splatting implementation by Kerbl~et~al.~\cite{kerbl3Dgaussians}, not all of them implement optimized viewers for inference.
As such, we intentionally avoid basing our benchmarks on these real-time viewers and instead came up with a way of removing most of the overhead caused by the PyTorch-based framework used for training and quality metric computation.
Specifically, we apply a ``baking'' step before rendering, which integrates all activation functions into the splat parameters and ensures all spherical harmonics coefficients are stored in the same buffer.
As shown in \Cref{tab:benchmarking_optimization}, this has a significant influence on the resulting frame rates of 3DGS and StopThePop, as they are highly optimized for real-time rendering, but a negligible influence for 2DGS and GOF.
Seeing that the obtained numbers align with what was shown in the supplemental video from 3DGS, we think that this is a fair method for comparison, which does not rely on method-specific optimizations for inference.
Note that in our implementation, the rendering is also controlled by a PyTorch-based framework.
Although not included in \cref{tab:benchmarking_optimization}, we also applied these these optimizations when measuring frame rates of StopThePop with opacity decay~\cite{radl2024stopthepop} and the approach of Huang~et~al.~\cite{huang2024erroranalysis}.

\begin{table}[bht]
\centering
\setlength\tabcolsep{5.76pt}
\small
\begin{tabular}{@{}lcccc}
\toprule
                                     & \multicolumn{2}{c}{Mip-NeRF360} & \multicolumn{2}{c}{Tanks\&Temples} \\
Method                               & Default & Optimized & Default & Optimized \\
\midrule
3DGS~\cite{kerbl3Dgaussians}         & 197.2   & 316.6     & 178.4   & 216.2     \\
StopThePop~\cite{radl2024stopthepop} & 157.3   & 212.5     & 180.6   & 206.2     \\
2DGS~\cite{huang20242d}            & 122.2   & 134.9     & 94.9    & 95.8      \\
GOF~\cite{yu2024gof}                 & 16.2    & 16.7      & 12.2    & 12.3      \\
\bottomrule
\end{tabular}
\caption{%
For fair comparison of frame rates, we ``bake'' activation functions into splat parameters and ensure all spherical harmonics coefficients are stored in the same buffer before rendering.
}\label{tab:benchmarking_optimization}
\end{table}

\section{Per-Scene Metrics}
\label{app_sec:scene_metrics}
We report per-scene SSIM, PSNR, and LPIPS for all evaluated scenes from Mip-NeRF360~\cite{barron2022mipnerf360} as well as Tanks and Temples~\cite{Knapitsch2017} in \Cref{tab:per_scene_metrics_m360} and \Cref{tab:per_scene_metrics_tt} respectively.
\begin{table*}[htb]
\centering
\setlength\tabcolsep{12.8pt}
{\scriptsize
\begin{tabular}{@{}lccccc|cccc|c}
\toprule
\multicolumn{11}{c}{SSIM$\uparrow$ on Mip-NeRF360~\cite{barron2022mipnerf360}\vspace{2pt}}\\
Method & \textit{Bicycle} & \textit{Flowers} & \textit{Garden} & \textit{Stump} & \textit{Treehill} & \textit{Bonsai} & \textit{Counter} & \textit{Kitchen} & \textit{Room} & \textit{Average} \\
\midrule
3DGS~\cite{kerbl3Dgaussians}         &   \cellcolor{3rd}0.770   &   \cellcolor{3rd}0.602   &  \cellcolor{1st}0.869   &  \cellcolor{3rd}0.774  &   \cellcolor{2nd}0.637    &  \cellcolor{3rd}0.938   &   \cellcolor{2nd}0.905   &   0.921   & 0.913  & \cellcolor{3rd}0.814  \\
StopThePop~\cite{radl2024stopthepop} &   0.767   &   0.600   &  0.865   &  \cellcolor{3rd}0.774  &   \cellcolor{3rd}0.634    &  \cellcolor{2nd}0.941   &   \cellcolor{2nd}0.905   &   \cellcolor{2nd}0.925   & \cellcolor{2nd}0.918  & \cellcolor{3rd}0.814  \\
~$\hookrightarrow$ w$\slash$ opacity decay &   0.757   &   0.592   &  0.858   &  0.773  &   0.629    &  \cellcolor{3rd}0.938   &   0.900   &   0.921   & \cellcolor{3rd}0.915  & 0.809  \\
2DGS~\cite{huang20242d}            &   0.732   &   0.570   &  0.841   &  0.756  &   0.615    &  0.931   &   0.891   &   0.915   & 0.906  & 0.795  \\
Huang~et~al.~\cite{huang2024erroranalysis}            &   0.764   &   0.601   &  0.865   &  0.771  &   0.632    &  \cellcolor{1st}0.942   &   \cellcolor{1st}0.907   &   \cellcolor{1st}0.926   & \cellcolor{2nd}0.918  & \cellcolor{3rd}0.814  \\
GOF~\cite{yu2024gof}                 &   \cellcolor{1st}0.787   &   \cellcolor{1st}0.634   &  \cellcolor{2nd}0.868   &  \cellcolor{1st}0.793  &   \cellcolor{1st}0.640    &  0.937   &  0.901   &   0.916   & 0.911  & \cellcolor{2nd}0.821  \\
Ours                                 &   \cellcolor{2nd}0.784   &   \cellcolor{2nd}0.629   &  \cellcolor{3rd}0.867   &  \cellcolor{2nd}0.792  &   \cellcolor{2nd}0.637    &  \cellcolor{2nd}0.941   &   \cellcolor{3rd}0.902   &   \cellcolor{3rd}0.923   & \cellcolor{1st}0.920  & \cellcolor{1st}0.822  \\
\midrule
Zip-NeRF~\cite{barron2023ICCV}       &   0.772   &   0.637   &  0.863   &  0.788  &   0.674    &  0.952   &   0.905   &   0.929   & 0.929  & 0.828  \\
\bottomrule
\end{tabular}
\begin{tabular}{@{}lccccc|cccc|c}
\toprule
\multicolumn{11}{c}{PSNR$\uparrow$ on Mip-NeRF360~\cite{barron2022mipnerf360}\vspace{2pt}}\\
Method & \textit{Bicycle} & \textit{Flowers} & \textit{Garden} & \textit{Stump} & \textit{Treehill} & \textit{Bonsai} & \textit{Counter} & \textit{Kitchen} & \textit{Room} & \textit{Average} \\
\midrule
3DGS~\cite{kerbl3Dgaussians}         &   \cellcolor{3rd}25.25   &   \cellcolor{2nd}21.52   &  \cellcolor{1st}27.41   &  26.55  &   \cellcolor{3rd}22.49    &  \cellcolor{2nd}31.98   &   \cellcolor{2nd}28.69   &   30.32   & 30.63  & 27.20  \\
StopThePop~\cite{radl2024stopthepop} &   25.18   &   \cellcolor{3rd}21.51   &  27.24   &  \cellcolor{3rd}26.63  &   \cellcolor{1st}22.51    &  \cellcolor{1st}32.00   &   \cellcolor{3rd}28.53   &   \cellcolor{2nd}31.09   & \cellcolor{2nd}30.98  & \cellcolor{2nd}27.30  \\
~$\hookrightarrow$ w$\slash$ opacity decay &   25.09   &   21.46   &  27.07   &  26.62  &   22.46    &  31.18   &   28.44   &   30.74   & \cellcolor{3rd}30.68  & 27.08  \\
2DGS~\cite{huang20242d}            &   24.74   &   21.10   &  26.63   &  26.20  &   22.33    &  31.22   &   28.09   &   30.29   & \cellcolor{3rd}30.68  & 26.81  \\
Huang~et~al.~\cite{huang2024erroranalysis}            &   25.19   &   21.45   &  \cellcolor{3rd}27.31   &  26.57  &   \cellcolor{2nd}22.50    &  \cellcolor{1st}32.00   &   \cellcolor{1st}29.01   &   \cellcolor{1st}31.31   & \cellcolor{1st}31.31  & \cellcolor{1st}27.41  \\
GOF~\cite{yu2024gof}                 &   \cellcolor{1st}25.46   &   \cellcolor{1st}21.69   &  \cellcolor{2nd}27.38   &  \cellcolor{1st}27.00  &   22.38    &  \cellcolor{3rd}31.57   &   \cellcolor{2nd}28.69   &   30.82   & 30.45  & \cellcolor{3rd}27.27  \\
Ours                                 &   \cellcolor{2nd}25.31   &   21.35   &  27.22   &  \cellcolor{2nd}26.85  &   22.37    &  31.55   &   28.40   &   \cellcolor{3rd}31.02   & 30.45  & 27.17  \\
\midrule
Zip-NeRF~\cite{barron2023ICCV}       &   25.85   &   22.33   &  28.22   &  27.35  &   23.95    &  34.79   &   29.12   &   32.36   & 33.04  & 28.56  \\
\bottomrule
\end{tabular}
\begin{tabular}{@{}lccccc|cccc|c}
\toprule
\multicolumn{11}{c}{LPIPS$\downarrow$ on Mip-NeRF360~\cite{barron2022mipnerf360}\vspace{2pt}}\\
Method & \textit{Bicycle} & \textit{Flowers} & \textit{Garden} & \textit{Stump} & \textit{Treehill} & \textit{Bonsai} & \textit{Counter} & \textit{Kitchen} & \textit{Room} & \textit{Average} \\
\midrule
3DGS~\cite{kerbl3Dgaussians}         &   \cellcolor{3rd}0.229   &   0.366   &  \cellcolor{1st}0.118   &  \cellcolor{3rd}0.244  &   0.367    &  0.253   &   \cellcolor{3rd}0.262   &   \cellcolor{3rd}0.158   & 0.289  & 0.254  \\
StopThePop~\cite{radl2024stopthepop} &   0.232   &   \cellcolor{3rd}0.365   &  \cellcolor{2nd}0.121   &  \cellcolor{3rd}0.244  &   \cellcolor{3rd}0.365    &  \cellcolor{3rd}0.248   &   \cellcolor{1st}0.254   &   \cellcolor{1st}0.154   & \cellcolor{2nd}0.280  & \cellcolor{3rd}0.252  \\
~$\hookrightarrow$ w$\slash$ opacity decay &   0.262   &   0.379   &  0.136   &  0.261  &   0.385    &  0.252   &   0.267   &   0.165   & 0.291  & 0.267  \\
2DGS~\cite{huang20242d}            &   0.301   &   0.404   &  0.165   &  0.300  &   0.434    &  0.280   &   0.291   &   0.180   & 0.316  & 0.297  \\
Huang~et~al.~\cite{huang2024erroranalysis}            &   0.238   &   0.368   &  0.124   &  0.249  &   0.376    &  0.253   &   \cellcolor{2nd}0.258   &   \cellcolor{2nd}0.156   & 0.288  & 0.257  \\
GOF~\cite{yu2024gof}                 &   \cellcolor{2nd}0.206   &   \cellcolor{1st}0.310   &  \cellcolor{3rd}0.122   &  \cellcolor{2nd}0.229  &   \cellcolor{2nd}0.326    &  \cellcolor{2nd}0.241   &   \cellcolor{2nd}0.258   &   0.166   & \cellcolor{3rd}0.286  & \cellcolor{2nd}0.238  \\
Ours                                 &  \cellcolor{1st} 0.204   &   \cellcolor{2nd}0.313   &  \cellcolor{3rd}0.122   &  \cellcolor{1st}0.226  &   \cellcolor{1st}0.321    &  \cellcolor{1st}0.235   &   \cellcolor{1st}0.254   &   0.159   & \cellcolor{1st}0.273  & 0.\cellcolor{1st}234  \\
\midrule
Zip-NeRF~\cite{barron2023ICCV}       &   0.228   &   0.309   &  0.127   &  0.236  &   0.281    &  0.196   &   0.223   &   0.134   & 0.238  & 0.219  \\
\bottomrule
\end{tabular}
}
\caption{
Per-scene image quality metrics for the Mip-NeRF360 dataset~\cite{barron2022mipnerf360} separated into outdoor and indoor scenes.
Excluding Zip-NeRF, the three best results are highlighted in \textcolor{1stText}{\textbf{green}} in descending order of saturation.
}\label{tab:per_scene_metrics_m360}%
\end{table*}

\begin{table*}[htb]
\centering
\setlength\tabcolsep{14.6pt}
{\scriptsize
\begin{tabular}{@{}lcccccccc|c}
\toprule
\multicolumn{10}{c}{SSIM$\uparrow$ on Tanks and Temples~\cite{Knapitsch2017}\vspace{2pt}}\\
Method & \textit{Family} & \textit{Francis} & \textit{Horse} & \textit{Lighthouse} & \textit{M60} & \textit{Panther} & \textit{Playground} & \textit{Train} & \textit{Average} \\
\midrule
3DGS~\cite{kerbl3Dgaussians} &  \cellcolor{3rd}0.871   &   \cellcolor{1st}0.901   &  0.889  &    \cellcolor{1st}0.834     & \cellcolor{1st}0.901 &   \cellcolor{1st}0.910   &    0.834     &  \cellcolor{2nd}0.791  & \cellcolor{1st}0.866  \\
StopThePop~\cite{radl2024stopthepop} &  \cellcolor{2nd}0.872   &   \cellcolor{3rd}0.899   &  0.889  &    \cellcolor{1st}0.834     & \cellcolor{1st}0.901 &   \cellcolor{1st}0.910   &    \cellcolor{3rd}0.835     &  \cellcolor{2nd}0.791  & \cellcolor{1st}0.866  \\
~$\hookrightarrow$ w$\slash$ opacity decay &  \cellcolor{3rd}0.871   &   0.897   &  \cellcolor{2nd}0.891  &    \cellcolor{2nd}0.833     & \cellcolor{2nd}0.896 &   \cellcolor{3rd}0.907   &    0.819     &  0.783  & \cellcolor{2nd}0.862  \\
2DGS~\cite{huang20242d}  &  0.860   &   0.884   &  0.883  &    0.820     & \cellcolor{3rd}0.887 &   0.902   &    0.806     &  0.768  & \cellcolor{3rd}0.851  \\
Huang~et~al.~\cite{huang2024erroranalysis}  &  \cellcolor{3rd}0.871   &   \cellcolor{2nd}0.900   &  \cellcolor{3rd}0.890  &    \cellcolor{1st}0.834     & \cellcolor{1st}0.901 &   \cellcolor{2nd}0.909   &    0.834     &  \cellcolor{3rd}0.790  & \cellcolor{1st}0.866  \\
GOF~\cite{yu2024gof} &  \cellcolor{1st}0.873   &   0.898   &  \cellcolor{2nd}0.891  &    0.825     & \cellcolor{1st}0.901 &   \cellcolor{2nd}0.909   &    \cellcolor{1st}0.840     &  \cellcolor{1st}0.793  & \cellcolor{1st}0.866  \\
Ours &  \cellcolor{2nd}0.872   &   0.896   &  \cellcolor{1st}0.893  &    \cellcolor{3rd}0.832     & \cellcolor{1st}0.901 &   \cellcolor{2nd}0.909   &    \cellcolor{2nd}0.839     &  0.786  & \cellcolor{1st}0.866  \\
\midrule
Zip-NeRF~\cite{barron2023ICCV} &  0.893   &   0.918   &  0.909  &    0.835     & 0.905 &   0.908   &    0.846     &  0.813  & 0.878  \\
\bottomrule
\end{tabular}
\begin{tabular}{@{}lcccccccc|c}
\toprule
\multicolumn{10}{c}{PSNR$\uparrow$ on Tanks and Temples~\cite{Knapitsch2017}\vspace{2pt}}\\
Method & \textit{Family} & \textit{Francis} & \textit{Horse} & \textit{Lighthouse} & \textit{M60} & \textit{Panther} & \textit{Playground} & \textit{Train} & \textit{Average} \\
\midrule
3DGS~\cite{kerbl3Dgaussians} &  \cellcolor{1st}25.05   &   \cellcolor{1st}27.64   &  \cellcolor{3rd}24.18  &    21.76     & \cellcolor{2nd}27.82 &   28.35   &    \cellcolor{1st}25.65     &  \cellcolor{1st}21.69  & \cellcolor{1st}25.27  \\
StopThePop~\cite{radl2024stopthepop} &  24.52   &   \cellcolor{3rd}26.91   &  23.88  &    21.65     & 27.70 &   \cellcolor{3rd}28.38   &    \cellcolor{3rd}25.49     &  \cellcolor{3rd}21.37  & 24.99  \\
~$\hookrightarrow$ w$\slash$ opacity decay &  24.64   &   26.69   &  \cellcolor{1st}24.40  &    \cellcolor{1st}22.32     & 27.42 &   28.18   &    24.91     &  21.25  & 24.98  \\
2DGS~\cite{huang20242d}  &  \cellcolor{3rd}24.72   &   26.29   &  23.86  &    21.49     & 26.89 &   27.94   &    24.29     &  20.89  & 24.55  \\
Huang~et~al.~\cite{huang2024erroranalysis}  &  \cellcolor{2nd}24.83   &   \cellcolor{2nd}27.59   &  24.13  &    \cellcolor{3rd}21.92     & \cellcolor{3rd}27.71 &   28.30   &    \cellcolor{2nd}25.55     &  \cellcolor{2nd}21.49  & \cellcolor{2nd}25.19  \\
GOF~\cite{yu2024gof} &  \cellcolor{1st}25.05   &   26.85   &  \cellcolor{2nd}24.30  &    21.17     & \cellcolor{1st}27.84 &   \cellcolor{1st}28.47   &    25.09     &  21.28  & \cellcolor{3rd}25.01  \\
Ours &  24.19   &   25.32   &  23.84  &    \cellcolor{2nd}21.96     & \cellcolor{2nd}27.82 &   \cellcolor{2nd}28.40   &    24.69     &  20.78  & 24.62  \\
\midrule
Zip-NeRF~\cite{barron2023ICCV} &  28.05   &   29.55   &  27.67  &    22.31     & 28.86 &   28.84   &    26.62     &  22.10  & 26.75  \\
\bottomrule
\end{tabular}
\begin{tabular}{@{}lcccccccc|c}
\toprule
\multicolumn{10}{c}{LPIPS$\downarrow$ on Tanks and Temples~\cite{Knapitsch2017}\vspace{2pt}}\\
Method & \textit{Family} & \textit{Francis} & \textit{Horse} & \textit{Lighthouse} & \textit{M60} & \textit{Panther} & \textit{Playground} & \textit{Train} & \textit{Average} \\
\midrule
3DGS~\cite{kerbl3Dgaussians} &  0.236   &   \cellcolor{2nd}0.344   &  0.239  &    \cellcolor{2nd}0.291     & 0.244 &   0.241   &    0.291     &  \cellcolor{2nd}0.320  & 0.276  \\
StopThePop~\cite{radl2024stopthepop} &  \cellcolor{3rd}0.228   &   \cellcolor{1st}0.338   &  \cellcolor{3rd}0.235  &    \cellcolor{1st}0.288     & \cellcolor{2nd}0.237 &   \cellcolor{2nd}0.236   &    \cellcolor{2nd}0.285     &  \cellcolor{1st}0.313  & \cellcolor{1st}0.270  \\
~$\hookrightarrow$ w$\slash$ opacity decay &  0.237   &   0.349   &  0.236  &    \cellcolor{3rd}0.302     & 0.255 &   0.248   &    0.318     &  \cellcolor{3rd}0.331  & 0.284  \\
2DGS~\cite{huang20242d}  &  0.272   &   0.386   &  0.264  &    0.333     & 0.278 &   0.266   &    0.349     &  0.366  & 0.314  \\
Huang~et~al.~\cite{huang2024erroranalysis}  &  0.234   &   \cellcolor{2nd}0.344   &  0.238  &    \cellcolor{2nd}0.291     & 0.245 &   0.241   &    0.291     &  \cellcolor{2nd}0.320  & 0.276  \\
GOF~\cite{yu2024gof} &  \cellcolor{1st}0.223   &   0.347   &  \cellcolor{2nd}0.230  &    0.310     & \cellcolor{3rd}0.238 &   \cellcolor{3rd}0.238   &    \cellcolor{3rd}0.286     &  \cellcolor{2nd}0.320  & \cellcolor{3rd}0.274  \\
Ours &  \cellcolor{2nd}0.227   &   \cellcolor{3rd}0.346   &  \cellcolor{1st}0.226  &    0.303     & \cellcolor{1st}0.233 &   \cellcolor{1st}0.235   &    \cellcolor{1st}0.276     &  \cellcolor{3rd}0.331  & \cellcolor{2nd}0.272  \\
\midrule
Zip-NeRF~\cite{barron2023ICCV} &  0.172   &   0.270   &  0.181  &    0.281     & 0.212 &   0.217   &    0.251     &  0.279  & 0.233  \\
\bottomrule
\end{tabular}

}
\caption{%
Per-scene image quality metrics for the Tanks and Temples dataset~\cite{Knapitsch2017}.
Excluding Zip-NeRF, the three best results are highlighted in \textcolor{1stText}{\textbf{green}} in descending order of saturation.
}\label{tab:per_scene_metrics_tt}
\end{table*}

\end{document}